# Expanding Click and Buy rates: Exploration of evaluation metrics that measure the impact of personalized recommendation engines on e-commerce platforms

[1]Namrata Chaudhary, [2]Drimik Roy Chowdhury

*Abstract:* To identify the most appropriate recommendation model for an e-commerce business, a live evaluation should be performed on the shopping website to measure the influence of personalization in real-time. The aim of this paper is to introduce and justify two new metrics — CTR-NoRepeat and Click & Buy rate — which stem from the standard metrics, Click-through(CTR) and Buy-through rate(BTR), respectively. The former variation tackles the issue of overestimation of clicks in the original CTR while the latter accounts for noting purchases of products that have been previously clicked, in order to validate that the buy included in the metric is a result of customer interactions. A significance test for independence of two means is conducted for multiple datasets, between each of the new metrics and its respective parent to determine the novelty and necessity of the variants. The Pearson-correlation coefficient is calculated to assess the strength of the linear relationships and conclude on the predictability factor amongst the aforementioned factors to investigate unknown connections between customer clicks and buys. Additionally, other metrics such as hits per customer, buyers per customer, clicks per customer etc. are introduced that help explain indicators of customer behavior on the e-commerce website in reference.

## I. INTRODUCTION

Consumer trends have seen a drastic change in the recent years with a shift towards online shopping. About 79% of U.S. consumers have switched to shopping online as stated by a 2016 TechCrunch report [1]. Thus, the market has been presented with numerous opportunities for online retailers to grow and capture greater revenue. This increase can be achieved by engaging these visitors to discover products that they are most likely to click or buy. Such products are generally shown on website widgets titled 'Recommended for you', 'You may also like' etc.. The re-ordering of products on these widgets can be achieved using a recommendation engine that uses past interaction data of individual customers along with contextual information like time, product attributes and promotions to predict the products a customer would click or purchase in the future.

There exists numerous algorithms that can be used to build a customized recommendation model, most of which fall under the three categories: collaborative, context-based and hybrid models. It is required for a business to identify which model is best-suited to predict recommendations for its e-commerce platform, and to determine this, one needs to evaluate these recommendation platforms live on the website. This practice is called online evaluation and the basic metrics used to measure the success of a recommendation model are Click-through rate (CTR), Add-to-Cart-through rate (ATC-TR) and Buy-through rate (BTR), which we define at the end of this section.

The purpose of this paper is to identify the shortcomings of these metrics when performing online evaluation on e-commerce websites, which are observed from specific shopping trends among online visitors that tend to inflate the basic metrics and therefore fail to reflect actual conversions resulting from recommended products. For example, on a jewelry based e-commerce website, we found that a customer would click on the same product multiple times in case of discounted products or during promotional days, which resulted in a CTR of about 20%, but only half of these clicks on recommended products were from unique products.

As a solution we explore an approach that defines new metrics that are variations of the standard CTR and BTR metrics — CTR-NoRepeat and Click & Buy rate — which can help overcome the aforementioned flaws and correctly reflect the success of a recommendation model on an e-commerce website. The outcomes of the independent samples t-test between the standard and newer CTR and BTR metrics confirm that the values reflected by the modified metrics brings forth previously unknown information that are shrouded behind the original metrics. Correlation between all these metrics is computed to ensure that they similarly reflect a recommendation model's ability to guide the customer journey and thus translate interactions into conversions. We also discuss other indicators like hits per customer, buyers per customer, clicks per customer and clicks per buy and elaborate on how they assess other significant aspects of customer behavior unreported from the online evaluation metrics on different online shopping platforms.

*Terms and Definitions:*

1. **CTR** is a ratio that shows how often people who see your products end up clicking it. It can be computed by dividing the number of times a recommended product was clicked by the number of times recommendations are seen. This metric is ideally reflective of the influence personalized recommendations have had on a customer's journey on the website.

2. **ATC-TR** and **BTR** are defined similarly as CTR but using add to carts and buys, respectively.

3. A web-component on a webpage of an e-commerce website displaying products suggested from a recommendation engine is a **widget**.

4. A list of products recommended to a user, placed in sequence on a widget will be referred to as a **Hit**.

In most cases, it is difficult to determine which recommendations displayed were actually noticed by a customer. For example, widgets containing products are not always directly visible on a page and the user might need to scroll or put in effort into seeing them. Therefore, we suggest introducing the restriction of a time limit of 5 minutes between a click and recommendation shown which can be extended to 30 minutes in case of add to carts and about 24 hours in case of a buys for the computation of CTR, ATC-TR and BTR, respectively. The reasoning behind this is based on a study by Kohavi R. [2] that states that an average customer spends 5 minutes navigating a website whereas a purchaser spends about 30 minutes. For the computation of BTR we extend this to 24 hours to allow customers considerable time to purchase a product. We found similar observations empirically where clicks made on products that were previously shown to a customer occurred within 5 minutes and add to carts within 30 minutes for approximately 90% of the cases.

## II. RELATED WORK

Beel J. and Langer S. in their paper [3] introduce modifications on the CTR metric as per the requirements that arise when evaluating Research-paper recommender systems. CTR, in their case, is defined as the ratio of clicks on recommended papers divided by the total recommended research papers, on which they introduce two variations:

i) CTR-Set is the mean of all individual CTRs for each search request. This metric helps average out the bias effect of multiple clicks that occur from a single search.

ii) CTR-User is a metric that calculates average CTR per user, thereby balancing the effect that a few highly active users might have.

iii) Additional metrics like Link-Through Rate, Annotate-Through Rate and Cite-Through Rate are also described in their paper in order to evaluate impact of each possible user action.

Similarly, Kim, Joon et al. in their paper [4], evaluate a news article recommendation model where they modify Click-through rate to identify the impact of certain words in the headlines of news articles by measuring the number of times headlines that were shown to the user contained a certain word to the number of times such headlines were clicked. It is computed as follows,

CTR of word w on day t := $wCTR(w,t) = clicks(w,t) / views(w,t)$,

where wCTR is the Click-through rate of a word present in the headlines Collins et al. introduce a validation test on the standard CTR metric in a study to analyze the role of position bias in information retrieval recommender systems [5]. In order to assess the bias that the position of a recommendation plays, placement of recommended products were shuffled and Click-through rates were calculated for each position of the recommendations' result. A positive correlation was observed between the precedence in a product's position and CTR, which led to the conclusion that customers tend to click on products placed first in line, independent of product relevance.

## III.  EVALUATION METRIC VARIATIONS

The evaluation metrics CTR, ATC-TR and BTR are designed to measure and examine the influence a recommendation model has on an online visitor of an e-commerce website to drive interactions with products, hopefully leading to a buy. However, an assumption we are making is that clicks, ATCs or buys that are included in their respective metrics are because of the influence of the product recommendations. This assumption is not always true and the calculation of these metrics is prone to certain shortcomings as listed below:

1. Online customers tend to view a product of their liking multiple times. As a result, the model assumes this product is important to the customer and might recommend it frequently. Such activity results in a higher CTR but these recurring clicks on the same item are not additional elements that should be considered in the CTR because of the effect of the recommended product being attributed already in the calculation. Therefore, the presence of such activity causes the CTR metric to overestimate the influence of recommendations on interactions.

2. While purchasing an item online, undecided customers may add and remove a product from the cart multiple times. Therefore, similar to the scenario in case(I), multiple add to carts on the same product by a customer inflates the ATC-TR metric.

3. For the calculation of BTR, the maximum time frame between a recommendation and a purchase of a product is set to be 24 hours in order to allow a customer to review his or her decision to buy. But in some cases, this 24 hour window allows for the possibility of purchases to be included that are not really a result of the recommendations' influence on the customer, thereby attributing more buys to BTR erroneously.

Hence, to overcome these possible errors, we introduce two other metrics, which are variations of the standard CTR and BTR: CTR - NoRepeat and Click & Buy rate. Moreover, we suggest adding a filter to the products recommended from the recommendation engine to refrain visitors from repeating actions on the same product and our model from displaying similar products recurrently. Ideally, this would enable customers to discover unforeseen relevant products of their liking.

Below are the variations we introduce in the two basic metrics -

1. **CTR - NoRepeat:** This is a variation of CTR where we eliminate all but the first click on a recommended product by a customer, given that the product has been clicked by the customer beforehand and attributed to a recommendation for the CTR calculation. This helps remove all repeated clicks and shows the true novelty factor [6] that a recommended product is able to bring in for a visitor. A similar concept can be applied for the calculation of an ATC-TR metric disregarding repeats.

2. **Click & Buy rate:** As discussed above, a shortcoming of the BTR metric is that it includes purchases within 24 hours of products from the time of the recommendation, but it is possible that the purchase was fallaciously ascribed to the recommendation and the customer might not have actually seen the widget. Therefore it is important to ensure that customers actually did interact with the recommendation widgets they are shown. To tackle this issue, an additional condition is checked in the computation of BTR, which is to inspect if the customer clicked on the recommended product within 5 minutes of the Hit timestamp before purchasing it in the next 24 hours. Note that this metric would not work for shopping sites that allow an option to directly add an item to cart before recording a click on it and viewing it on the display page. For such cases this metric could be modified to check if a customer has both added the product to cart within 30 minutes and then bought it within 24 hours of the recommendation timestamp.

There are two filters that can be applied to the resultant products of a recommendation model to prevent from showing a visitor repeated products:

1. Prevent the model from showing products to customers that have already been **clicked** by them *on that day*.

2. Prevent the model from showing products to customers that have already been **added to cart** by them in the *last seven days*.

With respect to the point regarding filtering products added to cart, an argument could be made both for and against this statement. This depends on whether a business wants to use recommended products as reminders for customers to purchase what they have previously added to cart (opted in case of grocery shopping sites), and in such cases, the ATC filter on recommendations should not be implemented. However, we advise that there is more value in using product recommendations as a means to show previously unclicked and not-added-to-cart products that customers have not yet had notable interactions with in order to show a greater variety of products.

## IV. SIGNIFICANCE TESTING

We have theoretically defined some variations of CTR and BTR in the previous section to overcome some unexpected scenarios that occur on e-commerce sites. However, it is also necessary to confirm that these metrics individually represent an essential KPI and are statistically different from one another. To deal with this we test for statistical difference between means of the metrics:

- CTR and CTR-NoRepeat
- BTR and Click & Buy rate

if $X_i$ denotes a Hit on the website where,

- Let $X_i$ = 1, if any product from the Hit was clicked/bought, with probability $p$
- Let $X_i$ = 0, if no product from the Hit was clicked/bought, with probability 1-$p$

The total click/buy activity $Y$ on $n$ Hits can be represented as the sum of the random variable $X_i$ as :

$Y = X_1 + X_2 + ... + X_n$

with a mean μ = np and variance $σ^2 = np(1−p)$. According to **Central Limit Theorem**, as each Hit is a bernoulli trial, so the sum/mean of $Y$ has a limiting normal distribution. That is, we can use normal distribution to approximate binomial distribution (sum of bernoulli) when n is large. The CTR and BTR metrics and their variations can be represented as a mean of $Y$, and we can use t-test for two samples to establish their independence. The mean and variance of $Y$ therefore can be represented as μ = p and $σ^2 = p(1−p)/n$, respectively.

One can use either pooled or unpooled t-test depending upon the ratio of the standard deviations of the two samples being compared. Pooled is used when this ratio is a factor between ½ and 2 and unpooled is used otherwise [7].

The corresponding hypotheses for the significance test between means of Metric1 and Metric2 would be

$H_0$: $\bar{Y}_{metric1}$ - $\bar{Y}_{metric2}$ = 0

$H_1$: $\bar{Y}_{metric1}$ - $\bar{Y}_{metric2}$ ≠ 0

We perform the t-test for an example data from an e-commerce website (*Case 1*), where based on the ratio of standard deviations between CTR and CTR-NoRepeat and BTR and Click & Buy rate, we find that a pooled test is appropriate [7].

Taking the level of significance **α** = 0.05, we reject $H_0$ if $p < 0.05$, where,

t test statistic = $\frac{(\bar{Y}1 - \bar{Y}2)}{s_p\sqrt{\frac{1}{n_1}+\frac{1}{n_2}}}$ and $s_p = \sqrt{\frac{(n_1-1)s_1^2+(n_2-1)s_2^2}{n_1+n_2-2}}$

for degrees of freedom := df = $n_1 + n_2 - 2$

$s_1^2 = \frac{p_1(1-p_1)}{n_1}$

$s_2^2 = \frac{p_2(1-p_2)}{n_2}$

### A. Test on CTR and CTR-NoRepeat

We start with observations from Case 1 collected over 11 days for the two metrics, CTR and CTR without repeats.

**Table 1: Pooled t-test for CTR and CTR-NoRepeat for Case 1**

|   | $\bar{Y}$ | Std. dev. | $n_1 = n_2$ | t-test statistic | p-value |
|---|---|---|---|---|---|
| **CTR** | 0.09 | 0.0012 | 103212 | 2663 | 0.0 |
| **CTR-NoRepeat** | 0.07 | 0.0011 | | | |

There is significant evidence to conclude that visitors on this e-commerce website performed repeated clicks on the same products over time. Thus, CTR-NoRepeat is a relevant metric that helps understand if the shown recommendations have aided customers in discovering pertinent products.

*B. Test on BTR and Click & Buy rate*

We use the same observations over 11 days and compute the two metrics, BTR and Clicks and Buy rate.

**Table 2: Pooled t-test for BTR and Click & Buy rate for Case 1**

|  | $\bar{Y}$ | Std. dev. | $n_1 = n_2$ | t-test statistic | p-value |
|---|---|---|---|---|---|
| **BTR** | 0.0042 | 0.00028 | 103212 | 1013 | 0.0 |
| **Click and Buy rate** | 0.0026 | 0.00022 |  |  |  |

Therefore, there is strong evidence once again on the premise that the two metrics are significantly different and that the Click & Buy rate represents a specific, distinct evaluation metric.

We performed the above process for 2 more e-commerce platforms and observed similar results in Tables 4 and 5 in the Appendix. The numbers for these calculations have been displayed additionally in the Appendix.

## V. PEARSON CORRELATION COEFFICIENT

We compute the Pearson correlation coefficient between the different evaluation metrics to get a sense of how they relate to each other. The results in Table 3 show that most metrics are positively-correlated and some specific insights are as follows:

1. A high correlation of 0.88 between CTR and CTR-NoRepeat was expected given that by definition the latter reflects a similar type of behavior as the former with the only additional requirement that repeated clicks be eliminated.

2. The moderately strong correlation factor of 0.77 between CTR and Click and Buy rate reflects that there is a linear relationship between recommender systems guiding a customer's journey by showing relevant products and these clicks resulting into purchases.

3. An important deduction is that the considerable difference of the correlation coefficients between BTR and CTR of 0.43 and BTR and CTR-NoRepeat of 0.68 signifies that the first click is a more accurate estimator of purchases than any subsequent repeated clicks on the same product.

**Table 3: Pearson correlation coefficient comparison across all metrics (Case 1)**

|  | CTR | CTR – NoRepeat | BTR | Click & Buy rate |
|---|---|---|---|---|
| **CTR** | 1.00 | 0.88 | 0.43 | 0.77 |
| **CTR - NoRepeat** | 0.88 | 1.00 | 0.68 | 0.90 |
| **BTR** | 0.43 | 0.68 | 1.00 | 0.69 |
| **Click & Buy rate** | 0.77 | 0.90 | 0.69 | 1.00 |

## VI. IMPLICATIONS

Identifying that CTR and CTR-NoRepeat are two significantly independent metrics implies that customers do perform repeated clicks on the same product in their journey. Moreover, with the stronger direct relationship identified between CTR-NoRepeat and BTR than that of CTR and BTR, a valid claim can be made in favor of using training data for a recommendation model after disregarding all repeated clicks of products for each customer within a single session. The cleaned data would better capture the relationship between consecutive clicks on different products by keeping only the first instance of a particular products' discovery for any customer.

The Click and Buy rate metric presents the correct value of conversions generated on an e-commerce website as a result of showing recommendations to customers. Without this additional metric present, the BTR in some cases, could present an inflated incremental conversion. However, it is the Click & Buy rate that informs the true increase in revenue which is crucial in determining which recommendation model to employ on any e-commerce website.

## VII.   ADDITIONAL METRICS

In addition to the CTR, BTR and their variations defined above, there are several other metrics that help us understand the traffic patterns of an e-commerce website and can potentially be a significant contributor to the success of the recommendation engine. These are:

1.  *Total hits per customer* indicates how many recommendations the average user receives on the e-commerce website. This metric infers to some degree the length of an average user session because to receive a large number of recommendations, the user must be hypothetically visiting multiple pages on the website. The length of the session can help indicate how many of the past interactions of a customer are relevant in order to predict the next recommended set of products. Note that for this metric to be valid, the recommendations must be implemented on a large proportion of the webpages of the e-commerce website.

2.  *Buyers per customer* demonstrates the unique number of buyers who received recommendations to the total number of unique customers. Often, this metric indicates conversions on a website and acts as an improved replacement to "buys per customer" when a large percentage of bulk buyers seem to be causing a massive skew in the aforementioned metric. Furthermore, bulk buyers are unlikely to be influenced by the recommendation engine as they probably know prior to their purchases, the products they intend to acquire.

3.  *Total clicks per customer* is simply the total number of clicks the average customer performs. One aspect of the user behavior that can be potentially revealed through this metric is the bounce rate — the percentage of users that leave the website only after a single page session. If the metric has a value defined near 1, then this serves as a strong indicator that a large proportion of the customers are not interacting with the website as intended. Perhaps a reason could be a monotonous and dull home page, dissuading users to proceed browsing on the e-commerce website.

4.  *Total clicks per buy* denotes how many clicks typically occur for a buy to take place. The metric under consideration moderately indicates the strength of the recommendation system through its ability to influence a user's purchasing decisions and to correctly display products for the customer's desires.

Although these metrics are simple in their calculation and idea, they individually and collectively analyze and evaluate the most important aspect of building the recommendation system on an e-commerce website: customer behavior. Once again, the data that would be used for the computation of the metrics defined above would be only inclusive of interactions of customers who received recommendations.

## VIII.   CONCLUSION

E-commerce platforms when powered with personalized recommendation systems can help increase customer clicks and purchases for the given online retailer. In order to decide which recommendation model is best suited for the particular e-commerce business, it is important to test various models and evaluate how customers interact with the presented recommendations, but the basic metrics used to do this might not always provide an accurate measure of success due to unpredictable behavior practiced by online visitors. To overcome such issues, this paper introduced two modifications of the original CTR and BTR metrics: CTR-NoRepeat and Click and Buy rate. These metrics were free from unnecessary repeated clicks or presence of false positives in the cases when customers were influenced by recommendations to purchase items. The paper further confirmed that validity of these metrics using significance and correlation testing. In addition, several other metrics are outlined that highlight specific attributes of customer behavior and can possibly showcase more interdependent relationships amongst customer interactions. This knowledge can serve as a road map for researchers who want to evaluate the true online success of their recommendation models.

# APPENDIX - A

Following are the results of significance testing calculation performed for datasets from two other different e-commerce businesses in Tables 4 and 5 to explore the validity of CTR-NoRepeat and Click and Buy rate metrics.

**Table 4: Pooled t-test for all metrics for Case 2**

|  | $\bar{Y}$ | Std. dev. | $n_1 = n_2$ | t | p-value |
|---|---|---|---|---|---|
| **CTR** | 0.08 | 0.0027 | 1905190 | 54535 | 0.0 |
| **CTR-NoRepeat** | 0.06 | 0.00024 | | | |
| **BTR** | 0.0035 | 6.06E-05 | 1905190 | 18301 | 0.0 |
| **Click and Buy rate** | 0.0021 | 4.69E-05 | | | |

**Table 5: Pooled t-test for all metrics for Case 3**

|  | $\bar{Y}$ | Std. dev. | $n_1 = n_2$ | t | p-value |
|---|---|---|---|---|---|
| **CTR** | 0.105 | 0.0009 | 227374 | 5829 | 0.0 |
| **CTR-NoRepeat** | 0.084 | 0.0008 | | | |
| **BTR** | 0.003 | 0.00016 | 227374 | 1794 | 0.0 |
| **Click and Buy rate** | 0.0019 | 0.00012 | | | |

It is observed that both the CTR-No Repeat and Click & Buy rate are significantly different from CTR and BTR, respectively, in both cases we tested.